\documentclass[10pt,twocolumn,letterpaper]{article}

\usepackage{cvpr}              

\usepackage{graphicx}
\usepackage{amsmath}
\usepackage{amssymb}
\usepackage{booktabs}
\usepackage{microtype}
\usepackage{algpseudocode}
\usepackage{algorithm}
\usepackage[numbers]{natbib}

\usepackage[pagebackref,breaklinks,colorlinks]{hyperref}

\newcommand{\algo}{{Note$_{EM}$}}

\usepackage[capitalize]{cleveref}
\crefname{section}{Sec.}{Secs.}
\Crefname{section}{Section}{Sections}
\Crefname{table}{Table}{Tables}
\crefname{table}{Tab.}{Tabs.}

\def\cvprPaperID{*****} 
\def\confName{CVPR}
\def\confYear{2022}

\begin{document}

\title{Unaligned Supervision for Automatic Music Transcription In-the-Wild}

\author{Ben Maman\\
Tel Aviv University\\
{\tt\small benmaman@mail.tau.ac.il}
\and
Amit H. Bermano\\
Tel Aviv University\\
{\tt\small amberman@tauex.tau.ac.il}
}
\maketitle

\begin{abstract}
   Multi-instrument Automatic Music Transcription (AMT), or the decoding of a musical recording into semantic musical content, is one of the holy grails of Music Information Retrieval. Current AMT approaches are restricted to piano and (some) guitar recordings, due to difficult data collection. In order to overcome data collection barriers, previous AMT approaches attempt to employ musical scores in the form of a digitized version of the same song or piece. The scores are typically aligned using audio features and strenuous human intervention to generate training labels.
We introduce \algo, a method for simultaneously training a transcriber and aligning the scores to their corresponding performances, in a fully-automated process. Using this \textit{unaligned supervision} scheme, complemented by pseudo-labels and pitch-shift augmentation, our method enables training on in-the-wild recordings with unprecedented accuracy and instrumental variety. Using only synthetic data and unaligned supervision, we report SOTA note-level accuracy of the MAPS dataset, and large favorable margins on cross-dataset evaluations. We also demonstrate robustness and ease of use; we report comparable results when training on a small, easily obtainable, self-collected dataset, and we propose alternative labeling to the MusicNet dataset, which we show to be more accurate. Our project page is available at \url{https://benadar293.github.io}.
\end{abstract}
\section{Introduction}
\label{section:introduction}
Automatic Music Transcription (AMT) is the task of decoding musical notes from an audio signal, and is one of the most central tasks in Music Information Retrieval (MIR). It benefits musicology and music education, musical search, and could even aid in realistic music synthesis.
AMT is challenging due to several reasons, such as notes sharing partial frequencies, polyphony (simultaneous notes played together, analogous to occlusions in computer vision), echo effects,
and multi-instrument performances, escalating complexity. 

Unsurprisingly, similarly to fields such as Computer Vision and Natural Language Processing, deep neural networks have contributed to AMT as well. However, as DNNs require massive amounts of training data, progress is limited. The main bottleneck is that manual annotation is severely infeasible, even if done by experts, as it requires highly precise timing. For this reason, for most instruments no datasets of highly accurate annotation have been collected. Collection efforts have concentrated mainly on two instruments. Guitar \cite{DBLP:conf/ismir/XiBPYB18, DBLP:conf/ismir/WigginsK19} annotations are done semi-automatically with human verification, in a difficult to scale process. For the piano, unique equipment (the Disklavier) logs key activity during performance, making annotation trivial and data collection simpler. Indeed, the guitar dataset we use for evaluation~\cite{DBLP:conf/ismir/XiBPYB18} (which is practically the only available one) consists of only $\sim$3 hours of recordings, compared to $\sim$140 hours of piano material~\cite{DBLP:conf/iclr/HawthorneSRSHDE19}. It is therefore not surprising that most AMT literature concentrates on the latter, where supervision and evaluation are clean and readily available~\cite{DBLP:conf/ismir/HawthorneESRSRE18, DBLP:conf/iclr/HawthorneSRSHDE19, DBLP:conf/ismir/HawthorneSSME21}. 

As it turns out, even within the case of the piano, supervised detectors struggle to generalize to variances in the instrument or environment, let alone from synthetic to real data. For this reason, for example, the accuracy of SOTA methods degrades in cross-dataset evaluation~\cite{DBLP:journals/corr/abs-2111-03017} (e.g., training on the piano recordings of the MAESTRO dataset~\cite{DBLP:conf/iclr/HawthorneSRSHDE19}, and testing on those of MAPS~\cite{unknown}).
To mitigate these data intensive requirements, a popular approach seeks to annotate existing recordings through alignment of real performances to their corresponding musical score. In other words, an easily obtainable digitized performance (or MIDI) of a musical piece is aligned to a real recorded performance. After the MIDI is warped to best match the recording, it is used as annotation. This is how, for example, the well known MusicNet dataset was constructed (with the support of human verification)~\cite{DBLP:conf/iclr/ThickstunHK17}. While promising, the alignment quality this approach demonstrates is not high enough to be used as labeling for network training. Indeed, the aforementioned dataset is notorious for its labeling inaccuracies~\cite{DBLP:conf/ismir/HawthorneESRSRE18, DBLP:journals/corr/abs-2111-03017}. 

In this work, we observe that the alignment process could be intertwined with the training of the transcriber, through the \textit{Expectation Maximization} (EM) framework. 
We introduce \algo, a framework that supports \textit{unaligned supervision}, based on easy-to-obtain musical scores to supervise in-the-wild recordings. The process comprises three steps (see Figure \ref{fig:overview}): first, we take an off-the-shelf architecture proposed for transcription, and bootstrap its training on synthetic data. Second, for the E-step, we use the resulting network to predict the transcription of unlabeled recordings. The unaligned score is then warped based on the predictions as likelihood terms, and used as labeling. For the M-step, the transcriber itself is trained on the new generated labels. Depending on the metric, best results were obtained when performing one or two such E-M iterations. In any case, alignment based on network predicted likelihoods is considerably more accurate than alignment based on spectral features~\cite{DBLP:conf/iclr/ThickstunHK17} (see Section~\ref{section:experiments}). It also enables better handling of inconsistencies between the audio and the score, which are inevitable.

Using this scheme, we achieve transcription accuracy that outperforms all existing methods on cross-dataset evaluations by a large margin for both the note- and frame-level metrics. For example, we reach 89.7\% note-level and 77.0\% frame-level F1 score on the MAESTRO test set (without using MAESTRO training data), where ~\citet{DBLP:journals/corr/abs-2111-03017} reach 28\% and 60\% when not including MAESTRO in the train set. 
Furthermore, we report note-level accuracy that compares or even surpasses fully supervised piano/guitar-specific transcription methods.
This is despite our method being trained on synthetic data and unaligned supervision alone. 

\algo also enables simple and convenient training on different instruments and genres. To demonstrate this, we train our network on other instruments, such as violin, clarinet, harpsichord, and many others - between 11-22 instruments, depending on the configuration. Furthermore, to evaluate the method's usability, we train it using a small-scale self-collected set of musical performances and corresponding unaligned supervision, and observe similar accuracy. We even generate alternative labeling to the aforementioned MusicNet dataset, which we denote MusicNet$_{EM}$, and demonstrate it is more accurate.  Finally, we also witness satisfying generalization capabilities, through the high quality transcription of unseen instruments and genres such as rock or pop (in which case transcription is pitch only).

Our contributions are as follows:
\begin{itemize}
    \item \algo -- A general framework for training polyphonic (multi-instrument) transcribers using unaligned supervision, allowing the use of in-the-wild recordings for training.
    
    \item A new SOTA note-level F1-score on the MAPS dataset of $87.3\%$ (vs. $86.4\%$ of supervised~\cite{DBLP:conf/iclr/HawthorneSRSHDE19}), and considerable improvement for cross-dataset evaluations. This is even though training is done using less supervision and less data ($\sim$34 vs. $\sim140$ hours). 
    
    \item unprecedented generalization to unseen instruments and musical genres. Results on these genres are unfortunately only qualitative due to lack of ground truth, but they are unmistakably favorable non-the-less.
    
    

    \item Alternative annotation for MusicNet, denoted MusicNet$_{EM}$, which is shown to be more accurate. 

\end{itemize}

\section{Related Work}

The two common forms of transcription are \textit{note-level}, where start (\textit{onset}) / end (\textit{offset}) note events are detected, 
and \textit{frame-level} transcription, where pitches are predicted at every given time,  implicitly determining the duration of notes. 
Other forms of transcription include stream-level, where the performance is segmented into different streams or voices. 
Segmentation can be according to instrument~\cite{DBLP:journals/taslp/WuC020, DBLP:journals/corr/abs-2111-03017}, but can also be between instances of the same instrument. 

While early works reduced the task of transcription to detection of active notes per-frame, later works~\cite{DBLP:conf/ismir/HawthorneESRSRE18, DBLP:conf/iclr/HawthorneSRSHDE19, DBLP:journals/taslp/WuC020} show the advantage of breaking down the detection into two components: onsets - beginning of notes, and frames - presence of notes. This is based on the observation that the more important and distinguished part of a note event is its onset. 


In multi-instrument transcription, the simpler form ignores instrument classes, assigning a single class for each pitch~\cite{DBLP:conf/icassp/WuCS19, DBLP:conf/mm/CheukHS21}. Only a handful of works also address, as we do, the problem of note-with-instrument transcription 
\cite{DBLP:journals/taslp/WuC020, DBLP:conf/icassp/ManilowSP20, DBLP:journals/corr/abs-2111-03017}. As we demonstrate (Section~\ref{section:experiments}), our approach provides cleaner and more attainable labeling, thus clearly surpassing the performance of these works. 

For piano transcription, the main benchmarks are MAPS~\cite{unknown} and MAESTRO~\cite{DBLP:conf/iclr/HawthorneSRSHDE19}. The MAPS dataset consists of synthetic and real piano performances, where usually the real performances are used for testing. MAESTRO is a large-scale dataset containing 140 hours of classical western piano performances, with fine and accurate annotation, generated using a Disklavier. The accurate annotation allows outstanding transcription quality~\cite{DBLP:conf/iclr/HawthorneSRSHDE19, DBLP:conf/ismir/HawthorneSSME21, DBLP:journals/corr/abs-2111-03017}. However, the main drawback of this dataset is the lack of variety: It contains only piano recordings, which prevents generalization to other musical instruments, and even to varieties in recording environments and pianos. Thus, transcription quality degrades significantly even when testing the model on other piano test sets, such as MAPS.

For annotation of guitar transcription, \citet{DBLP:conf/ismir/XiBPYB18} rely on hexaphonic pickup (separated to 6 strings), breaking the problem down into annotation of monophonic music which is simpler than polyphonic. 
Unfortunately, this approach still requires manual labor, which limits broad data collection. This results in a small dataset - 3 hours in total. Hence, this dataset can be used for evaluation but is less effective for training in-the-wild transcribers.

For other instruments, or multi-instrument transcription, the main existing dataset is MusicNet~\cite{DBLP:conf/iclr/ThickstunHK17}, which contains 34 hours of classical western music, performed on various instruments. The annotation was obtained by aligning separate-sourced (i.e. by other performers) MIDI performances, rendered into audio, with the real recordings, according to low frequencies. This dataset has the clear advantage of variety, both in instruments and in recording environments, as recordings were gathered from many different sources. However, despite being verified by musicians, the alignment is of poor quality, and timing of notes is not precise, significantly inhibiting learning and performance, as we show. Similar datasets exist - SU~\cite{DBLP:conf/cmmr/SuY15}, extended SU~\cite{DBLP:journals/taslp/WuC020}, and URMP~\cite{URMP} datasets, which suffer from similar limitations and are small.


On the task of instrument-sensitive transcription (note-with-instrument), few works have been done, because 
of the aforementioned limitations of multi-instrument datasets.~\citet{DBLP:journals/taslp/WuC020} train and test on MusicNet for this task, but reported note-level accuracies are very low, below 51\% on all instruments except for piano and violin, on which the accuracies are $\sim$69\% and $\sim$61\% respectively.~\citet{DBLP:journals/corr/abs-2111-03017} train on a mixture of datasets - MAESTRO, GuitarSet, MusicNet and Slakh2100 (Synthetic). They map the spectrogram into a sequence of semantic midi events, taking an NLP seq2seq approach. This setting is flexible and allows to easily represent multi-instrument transcription. However, the performance on the cross-dataset, or zero-shot task, is low (below 33\% on note-level F1), and performance on MusicNet is low, even when training on MusicNet (50\% note-level F1 at most). 


It is important to note, that none of the latter works propose any framework or method for weakly- or self-supervised transcription. \citet{DBLP:conf/mm/CheukHS21} train instrument-insensitive transcription without supervision using a reconstruction loss and Virtual Adversarial Training~\cite{DBLP:journals/pami/MiyatoMKI19}, but as we show, our framework performs much better, and also allows instrument-sensitive transcription. To our knowledge, our work is the first to propose such a framework for multi-instrument polyphonic music, including instrument-sensitive transcription. 




\section{Method}\label{section:method}
\begin{algorithm}[tb]
   \caption{Transcription EM}
   \label{alg:em}
\begin{algorithmic}
  \State {\bfseries Input:} audio $a_1, \dots a_N$, unaligned MIDI $m_1,\dots m_N$
   \State {\bfseries Output:} transcriber $f_\Theta$, labels $y_1, \dots y_N$
   \State pre-train $f_\Theta$ (synthetic)
   \State $y_i, d_i = None, \infty\quad i=1,\dots,N$
   \Repeat
   \For{$i=1$ {\bfseries to} $N$}
   \State $y_i^{temp}, d_i^{temp}= DTW(f_\Theta(a_i), m_i)$
   \If {$d_i^{temp} < d_i$}
   \State $y_i, d_i = y_i^{temp}, d_i^{temp}$
   \EndIf
   \EndFor
   \State $\Theta = argmin\frac{1}{N}\sum_{i=1}^N L(f_\Theta, a_i, y_i)$
   \Until{$\frac{1}{N}\sum_{i=1}^N d_i$ converges}
\State \textbf{return} $f_\Theta$, $y_1, \dots y_N$
\end{algorithmic}
\end{algorithm}




\begin{figure*}
\includegraphics[width=0.99\textwidth]{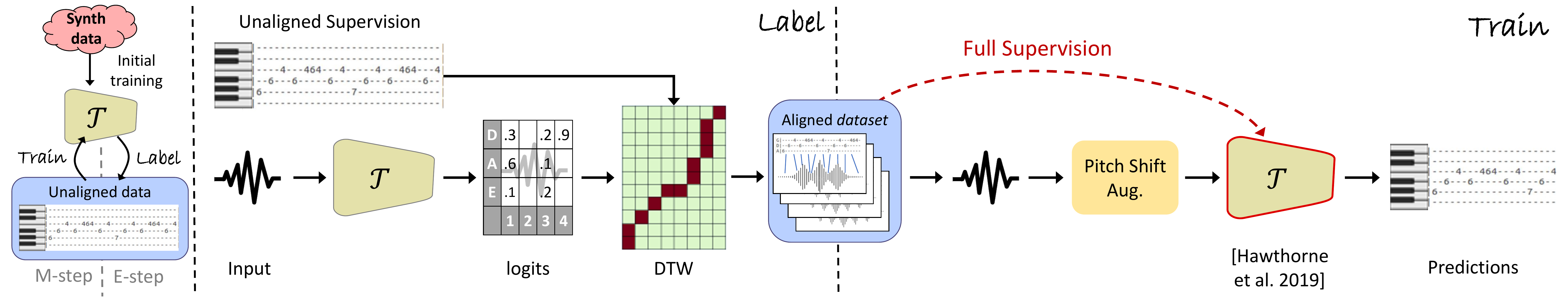}
\caption{\algo system overview. Left: the overall EM approach. Given a synthetic or otherwise supervised dataset, and an unaligned domain, we start by training the transcriber $\mathcal{T}$ on the synthetic data. Next, we use the transcriber to \textit{label} the domain (E-step, middle). We use this as supervision for further training, resulting in a stronger $\mathcal{T}$ model (M-step, right). Middle: At the core of our unaligned supervision scheme is the alignment step. Probabilities for each note at each timestep are computed using $\mathcal{T}$. Then, the unaligned labels are warped using DTW to maximize said logits. Right: the warped results are accumulated into the aligned dataset, which can be used to retrain $\mathcal{T}$. During training we use pitch shift augmentation, to improve robustness and performance.}
\label{fig:overview}
\end{figure*}

The key observation of our method is that a weak transcriber can still produce accurate predictions if the global content of the outcome is known up to a warping function. These accurate predictions, in turn, can be used as labels to further improve the transcriber itself. As we demonstrate (see Section~\ref{section:experiments}), this approach is more accurate than that of pseudo-labels (see Section~\ref{subsection:pseudo}), due to the unaligned known global content. The weak transcriber thus transforms weak supervision into full supervision and refines itself.




Our method, described in pseudo-code Algorithm~\ref{alg:em}, relies on \textit{Expectation Maximization (EM)} (see Section~\ref{sec:em}), and involves three components (see Figure~\ref{fig:overview} left): (I) Synthetic data initial training (Section ~\ref{section:synth}), (II) aligning real recordings with separate-source MIDI (Section~\ref{subsection:alignment}), including deciding which frames to use and which not to (Section~\ref{subsection:pseudo}). (III) transcriber refinement, including pitch-shift equivariance augmentations (Section~\ref{equivariance}).

\subsection{Expectation Maximization (EM)}
\label{sec:em}
Expectation Maximization (EM) is a paradigm for unsupervised or weakly-supervised learning, where labels are unknown, and are assigned according to maximum likelihood. It can be formulated as an optimization problem:
\[
\Theta^* = \underset{\Theta}{argmax}\;\underset{y_1,\dots, y_n}{max}\;P_\Theta(a_1, \dots, a_n, y_1, \dots, y_n)
\]
where $a_1,\dots,a_n$ are data samples, and $y_1,\dots,y_n$ are their unknown labels. The optimization problem can be solved by alternating steps, repeated iteratively until convergence (assuming some pre-training or bootstrapping of $\Theta$): 
\begin{align}
&y_1,\dots,y_n = \underset{y_1,\dots,y_n}{argmax}\; P_\Theta(a_1, \dots, a_n, y_1, \dots, y_n)\label{line1}\\
&\Theta = \underset{\Theta}{argmax}\;P_\Theta(a_1, \dots, a_n, y_1, \dots, y_n)\label{line2}
\end{align}
which are referred to as the \textit{E-step} (\ref{line1}) and the \textit{M-step} (\ref{line2}). 

In our scenario, the data samples $a_1, \dots, a_n$ are the unlabelled audio recordings, and $y_1, \dots, y_n$ are the unknown per-frame labels. We assume that the recordings are performances of pre-defined musical pieces $m_1, \dots, m_n$, such as in classical music, in the form of MIDI from other performers. We perform the E-step by aligning $m_1, \dots, m_n$ with the predicted probabilities over $a_1, \dots, a_n$ using dynamic time warping (DTW)~\cite{dtw_cite}.
We initialize $\Theta$ by training on synthetic data which is (trivially) supervised.



\subsection{Initial training}
\label{section:synth} 

We use synthetic data (see Section~\ref{sec:data} for details) to train the architecture proposed by~\citet{DBLP:conf/iclr/HawthorneSRSHDE19}. Of course, our training scheme can also be applied to other architectures, but this one has proven to be effective for supervised piano transcription, reaching 95\% note-level and 90\% frame-level F1 scores. It has separate detection heads for onsets, offsets, and frames, allowing to perform alignment according to semantic information. As we show (see Supplementary), onset information is the most effective for alignment. 
This initial network is trained to detect only pitch, without instrument, but it can also be further trained to detect instrument as well (see section~\ref{subsubsection:instrument_sensitive}).

\subsection{Labeling} 
We label real data using dynamic time warping between the initial network's predicted probabilities and the corresponding MIDIs. This is contrary to \cite{DBLP:conf/iclr/ThickstunHK17}, who compute the dynamic time warping in the frequency space. As can be seen in the Supplementary, MIDI guided alignment yields more accurate labels than simple thresholding. It also provides instrument information.

The alignment process is depicted in Figure~\ref{fig:overview} middle, and essentially relies on Dynamic Time Warping. Using DTW, we search for a chronologically monotonic mapping between the unaligned labeling and its corresponding recording, such that for each selected note the probability, as predicted by the transcription model, is maximized.



We argue that using the network's predicted probabilities as local descriptors for DTW 
has the following advantages:

\textbf{(i) Inconsistencies} -- For a separate-source MIDI (i.e., originating from a different performer), inconsistencies between the performances in inevitable. This includes repetitions of cadenzas, and more subtle nuances, such as trills, or in-chord order changing. Precise onset timing can be adjusted locally for each note independently according to predicted likelihoods. Failed detection, whether false positive or false negative, can be avoided based on network's probabilities, i.e., pseudo-labels can also be leveraged in addition to the alignment. 

\textbf{(ii) Label refinement} - the labeling process can be repeated during training, thus refining the labels, since the network has improved. 

\textbf{(iii) DTW computation speed} - for DTW descriptors, we project the 88 pitches into a single octave (12 pitches) using maximum activation across octave, hence representation length for DTW is 12 rather than 50~\cite{DBLP:conf/iclr/ThickstunHK17}. This has an impact on computation speed because DTW requires quadratic time. After projection, for an audio recording of $\sim$2:30 minutes, DTW takes $\sim$1 second. 

\paragraph{Pseudo Labels}\label{subsection:pseudo}
As aforementioned, the alignment can produce false detections, whether positive or negative. 
To avoid this false detection automatically, and still leverage all data, we label classes with predicted confidence above a threshold $T_{pos}$ as positive, and classes with predicted confidence beneath a threshold $T_{neg}$ as negative, regardless of the alignment. Classes with probability $0.5<p<T_{pos}$ which were not marked positive are considered unknown and we do not back-propagate loss through them. We do this to allow detection of onsets undetected by the labeling. We do not do the same for negative detection (i.e., $T_{neg}<p<0.5$) as there is already a strong bias against onset detection, as onsets are very sparse (an onset lasts a single frame).

In our experiments we use thresholds $T_{pos} = 0.75$ and $T_{neg} = 0.01$ for all classes - onsets, frames and offsets. We can use a low negative threshold since the MIDI performance already constrains the labels, and activations (whether onset, frame, or offset) are sparse, thus mode collapse is less of an issue. 

\subsection{Tonality - Pitch Shift Equivariance}\label{equivariance}
Music transcription has a unique inherent structure, where a pitch shift on the waveform induces a corresponding predetermined translation of the labels. 
We leverage this structure by enforcing consistency across pitch shift: We create 11 additional pitch shifted copies of our data, with pitch shifts (in semitones):
$s_i~=~i~+~\alpha_i,\; -5~\le~i~\le~5,\;  \alpha_i~\sim~\mathcal{U}(-0.1, 0.1)$,
where $\mathcal{U}(-0.1, 0.1)$ is the uniform distribution on the interval $[0,1]$, as suggested by Thickstun et al.~\cite{DBLP:conf/icassp/ThickstunHFK18}. \textbf{We compute the labels only for the original copy}, and for each copy shift labels accordingly. This not only augments the data by an order of magnitude, but also implicitly enforces consistency across pitch shift, serving as a regularization, forcing the model to learn tonality.

\subsection{Instrument-Sensitive Transcription (note-with-instrument)}
In this setting, we define a distinct class for each combination of pitch and instrument, i.e., the number of classes $C$ is (number of pitches)$\cdot$(number of instruments). 

We start with instrument-insensitive training on synthetic data. To adjust the transcriber to the new task of detecting also instrument, we duplicate the weights of the final linear layer of the onset stack $I + 1$ times: once for each instrument, and one copy to maintain instrument-insensitive prediction. This redundancy serves as regularization and improves learning. Thus, at the beginning of instrument-sensitive training, upon detection of a note, the transcriber will detect the note as active on all instruments. During training the transcriber will learn to separate instruments, according to the labels. We apply the same labelling process to this scenario as well - the difference only being more classes. We maintain the low representation length of $12$ for DTW computation by maximizing activation both across octave and instrument. To allow the transcriber (which is initially insensitive to instrument) to learn instrument separation, we do not use pseudo-labels in the initial labelling, only from the second labelling iteration.

\section{Experiments}\label{section:experiments}
\begin{table*}
\caption{Piano transcription results (Precision, Recall, and F1 scores).~\citet{DBLP:journals/corr/abs-2111-03017} was trained for instrument-sensitive transcription. Notice the drop in performance when excluding MAESTRO from training in the zero-shot task (ZS). The rest in an instrument-insensitive setting. 
'Synth' is trained only on synthetic data, and is the result of our initial training step. All following models are fine-tuned from it:
'MusicNet' is fine-tuned on the MusicNet annotation. Notice performance reduction compared to Synth, indicating low quality labeling. 'MusicNet$_{EM}$' is fine-tuned on our annotation, with two labeling iterations. 'MusicNet$_{EM}$1L' is with a single labeling iteration, and 'self-collected' is using $\sim$30 hours of piano and guitar recordings, with our annotation. 
As can be seen, our approach surpasses fully supervised note-level accuracy on the MAPS test set, and is comparable for MAESTRO. The presented pitch augmentation's effect is evaluated by adding it to 'MusicNet' training, and removing it from 'MusicNet$_{EM}$'.
}\label{transcription_table}
\begin{center}
\begin{tabular}{|c|c|c|c|c|c|c|c|c|c|c|c|c|}
\hline
 & \multicolumn{6}{|c|}{MAESTRO} & \multicolumn{6}{|c|}{MAPS} \\
\hline
 & \multicolumn{3}{|c|}{Note} & \multicolumn{3}{|c|}{Frame} & \multicolumn{3}{|c|}{Note} & \multicolumn{3}{|c|}{Frame} \\
\hline
\textbf{Supervised} & P & R & F1 & P & R & F1 & P & R & F1 & P & R & F1 \\
\hline
~\citet{DBLP:conf/iclr/HawthorneSRSHDE19} & \textbf{98.3} & 92.6 & 95.3 & \textbf{92.1} & 88.4 & \textbf{90.2} & - & - & \textbf{86.4} & - & - & \textbf{84.9} \\
~\citet{DBLP:journals/taslp/KongLSWW21} & 98.2 & \textbf{95.4} & \textbf{96.7} & 88.7 & \textbf{90.7} & 89.6 &- & - & - & - & - & - \\
~\citet{DBLP:journals/corr/abs-2111-03017} & - & - & 96.0 & - & - & 88.0 & - & - & - & - & - & - \\
\hline
\textbf{Weakly/self-supervised} &\multicolumn{12}{|c|}{} \\
\hline
~\citet{DBLP:journals/corr/abs-2111-03017} ZS & - & - & 28.0 & - & - & 60.0 & - & - & - & - & - & - \\
~\citet{DBLP:conf/mm/CheukHS21} & - & - & - & - & - & - & 86.1 & 67.3 & 75.2 & \textbf{88.8} & 72.7 & 79.5\\
\hline
Synth & 86.0 & 82.1 & 83.8 & 79.1 & 72.6 & 74.7 & 79.5 & 79.3 & 79.1 & 85.0 & 70.9 & 76.6 \\
\hline
MusicNet & 59.3 & 43.2 & 49.7 & 78.8 & 55.0 & 62.0 & 54.8 & 43.3 & 48.1 & 69.2 & 75.5 & 71.4 \\
MusicNet (pitch aug.) & 68.1 & 50.3 & 57.5 & \textbf{81.6} & 48.8 & 57.9 &  59.0 & 49.1 & 53.4 & 71.2 & 79.9 & 74.3 \\
\hline
MusicNet$_{EM}$ (ours) & 92.6 & \textbf{87.2} & \textbf{89.7} & 77.4 & 76.1 & 76.0 & 88.2 & \textbf{86.5} & \textbf{87.3} & 84.4 & 76.7 & 79.6 \\
MusicNet$_{EM}$1L (ours) & \textbf{95.6} & 84.7 & \textbf{89.7} & 79.1 & 76.9 & \textbf{77.0} & \textbf{90.3} & 83.7 & 86.8 & 86.2 & 78.0 & \textbf{81.4} \\
Self-collected (ours) & 93.5 & 86.2 & 89.6 & 76.3 & \textbf{79.3} & 76.8 & 88.8 & 84.6 & 86.6 & 81.6 & \textbf{81.1} & 80.9 \\
w/o pitch aug. (ours) &91.1 & 85.6 & 88.1 & 76.3 & 74.8 & 74.3 & 85.9 & 83.7 & 84.7 & 83.9 & 74.0 & 78.0 \\
\hline
\end{tabular}
\end{center}
\end{table*}
\begin{table*}
\caption{
String and wind instruments separately. In the table we use the same test split as \citet{DBLP:conf/mm/CheukHS21} (excluding piano pieces, which are less reliable compared to MAPS and MAESTRO). 
Results compare the same training on three different datasets (rows), evaluated on both the given MusicNet annotations, and the ones generated using our unaligned supervision scheme (columns). 
We also compare against \citet{DBLP:journals/corr/abs-2111-03017}, which use a different split. 
}\label{table:musicnet_combined}
\begin{center}
\begin{tabular}{|c|c|c|c|c|c|c|c|c|c|c|c|c|}
\hline
& \multicolumn{3}{|c|}{Note} & \multicolumn{3}{|c|}{Frame} & \multicolumn{3}{|c|}{Note} & \multicolumn{3}{|c|}{Frame} \\
\hline
& P & R & F1 & P & R & F1 & P & R & F1 & P & R & F1 \\
\hline
& \multicolumn{6}{|c|}{MusicNet$_{EM}$ \textbf{Strings} test} & \multicolumn{6}{|c|}{MusicNet \textbf{Strings} test} \\
\hline
~\citet{DBLP:conf/mm/CheukHS21} &&&&&&& 63.6 & 58.8 & 61.0 & \textbf{78.9} & 60.7 & 68.4 \\
\hline
MusicNet & 36.0 & 33.3 & 34.6 & 58.5 & 69.4 & 63.4 & 44.1 & 37.1 & 39.9 & 66.2 & \textbf{73.5} & \textbf{69.4} \\
\hline
Synth & 73.0 & 59.7 & 65.2 & 70.3 & 45.5 & 54.4 & 57.3 & 44.2 & 49.1 & 66.7 & 40.9 & 49.8 \\
MusicNet$_{EM}$ (ours) & \textbf{81.8} & \textbf{78.7} & \textbf{80.0} & \textbf{73.2} & \textbf{69.8} & \textbf{71.3} & \textbf{68.6} & \textbf{61.1} & \textbf{63.9} & 72.4 & 65.0 & 68.3 \\
\hline
& \multicolumn{6}{|c|}{MusicNet$_{EM}$ \textbf{Wind} test} & \multicolumn{6}{|c|}{MusicNet \textbf{Wind} test} \\
\hline
~\citet{DBLP:conf/mm/CheukHS21} & &&&&&& 48.6 & 47.9 & 48.2 & 69.8 & 65.8 & 67.4 \\
\hline
MusicNet & 50.3 & 46.6 & 48.4 & 66.8 & 75.0 & 70. & 40.0 & 36.3 & 38.0 & 69.9 & 78.3 & 73.4 \\
\hline
Synth & 80.4 & 77.2 & 78.8 & \textbf{72.7} & 59.3 & 65.3 & 56.8 & 54.0 & 55.4 & \textbf{71.8} & 58.5 & 64.3\\
MusicNet$_{EM}$ & \textbf{84.2} & \textbf{91.1} & \textbf{87.5} & 71.4 & \textbf{79.0} & \textbf{75.0} & \textbf{58.9} & \textbf{63.1} & \textbf{60.9} & 70.7 & \textbf{78.2} & \textbf{74.2} \\
\hline
& \multicolumn{6}{|c|}{} & \multicolumn{6}{|c|}{~\citet{DBLP:journals/corr/abs-2111-03017} test split} \\
\hline
~\citet{DBLP:journals/corr/abs-2111-03017} &\multicolumn{6}{|c|}{} &- & - & 50.0 & - & - & 68.0 \\
\hline
\end{tabular}
\end{center}
\end{table*}
\begin{table}
\caption{Transcription results on GuitarSet. MusicNet$_{EM}$ is the MusicNet recordings with our annotation. Note-level metrics of  \citet{DBLP:conf/ismir/XiBPYB18} and \citet{DBLP:conf/ismir/WigginsK19} are unavailable.
Note that our results is for an \textbf{unseen} instrument, since MusicNet recordings contain no guitar performances.~\citet{DBLP:journals/corr/abs-2111-03017} reach high accuracy on GuitarSet when training on GuitarSet, but perform poorly when generalizing from one dataset to another, in the zero-shot task (ZS), where GuitarSet data is excluded from the train set.
}\label{guitarset_table}
\begin{center}
\begin{tabular}{|c|c|c|}
\hline
 & Note F1 & Frame F1 \\
\hline
\textbf{Supervised} & \multicolumn{2}{|c|}{} \\
\hline
~\citet{DBLP:conf/ismir/XiBPYB18} & - & 64.6 \\
~\citet{DBLP:conf/ismir/WigginsK19} & - & 82.6 \\
~\citet{DBLP:journals/corr/abs-2111-03017} & \textbf{90.0} & \textbf{89.0} \\
\hline
\textbf{Weakly/self-supervised} & \multicolumn{2}{|c|}{} \\
\hline
~\citet{DBLP:journals/corr/abs-2111-03017} ZS & 32.0 & 58.0 \\
\hline
Synth  & 68.4 & 72.9 \\
MusicNet orig. & 10.0 & 57.2 \\
MusicNet$_{EM}$ (ours) & \textbf{82.9} & \textbf{81.6} \\
Self-Collected (ours) & 82.2 & 79.3 \\
w/o pitch aug. (ours) & 75.4 & 77.8 \\
\hline
\end{tabular}
\end{center}
\end{table}

For all our experiments, we use an architecture similar to the one proposed by \citet{DBLP:conf/iclr/HawthorneSRSHDE19}, but wider, to handle variety in instruments: We use LSTM layers of size 384, convolutional filters of size 64/64/128, and linear layers of size 1024.

We re-sampled all recordings to $16kHz$ sample rate, and used the log-mel spectrogram with 229 bins as the input representation. We used hop length 512. We used the mean BCE loss, with an Adam optimizer, with gradient clipped to norm $3$, and batch size $8$. The initial synthetic model was trained for 
$350K$ 
steps. This took 65 hours on a pair of Nvidia GeForce RTX 2080 Ti GPUs. Further training on real data was done for $90 * |Dataset|$ steps.
In the case of MusicNet$_{EM}$, this is $\sim 90 * 310 = 28K$ iterations. For most experiments, labeling is performed twice: once after sythetic training, and once after $45 * |Dataset|$ steps. Training on MusicNet$_{EM}$, for 28K iterations including 2 labelling iterations which require DTW, took 16 hours on a pair of Nvidia GeForce RTX 2080 Ti GPUs.

In the following, we discuss the data we have used during our evaluations (Section~\ref{sec:data}), we report quantitative results (Section~\ref{sec:eval}), and compare to previous work (throughout the evaluations of Section~\ref{sec:eval}). The affects of the pitch-shift augmentations can be seen in Tables~\ref{transcription_table}, and~\ref{guitarset_table}. Further ablations studies, considering various steps, such as the pseudo-labeling, EM iterations, alignment quality, and others can be found in the supplementary material (Section~\ref{sec:ablations}). 
\subsection{Data \& Instrument Distribution}
\label{sec:data}
In our experiments, we use three datasets:

\textbf{MIDI Pop Dataset}~\cite{midiDataset} is a large collection of MIDI files. The data consists of almost $80,000$ songs, from which we used $\sim8,500$ randomly selected ones. $\sim4500$ of the performances, of length 278:09:01 hours, are \textit{mp3} compressed, and the rest with lossless \textit{flac} compression. In total 501:11:30 hours of audio were synthesized from MIDI. We use this dataset to bootstrap the process, by training the system to transcribe the rendered audio according to the original MIDI. Note that for flexibility, we only use pitch labels from this data, without instrument specific labels. \textbf{We use this dataset only for pre-training}.

\textbf{MusicNet}~\cite{DBLP:conf/iclr/ThickstunHK17} comprises recordings of multiple instruments in an unbalanced mix. Labels for this dataset are of lower quality, as they are generated by alignment to musical scores, but in preprocess. Most recordings are of a piano ($\sim$15 out of $\sim$ 34 hours are piano solo, and $\sim$7 other hours include the piano). We use the recordings of this dataset, and their provided unaligned corresponding musical scores.  Instead of the provided labels (or aligned scores), we offer MusicNet$_{EM}$ (in the supplementary material) -- alternative labeling generated by our framework -- and demonstrate their superiority (Section~\ref{section:experiments}).

Our \textbf{Self-Collected dataset} demonstrates the simplicity of collecting data for our method. We gather $74$ additional hours of recordings, 
including over $30$ hours of orchestra, $5$ hours of solo guitar (pieces by Albeniz, Sor, and Tarrega), $11$ hours of harpsichord ($6$ hours solo), and more. We use this data to supplement or replace MusicNet in our experiments.

Qualitative results in the accompanying video are from a model trained on all three datasets (the \textbf{MIDI Pop} dataset used only for pre-train).
Our generated annotation for MusicNet, and our code, together with qualitative examples for various genres and instruments, are available at \url{https://benadar293.github.io}.


\subsection{Evaluation}\label{sec:eval}
For our experiments,  
we train only on MusicNet$_{EM}$ and/or self collected in-the-wild data, where the model is pre-trained on synthetic data. \textbf{We do not use MAESTRO, MAPS, or GuitarSet for training}. We evaluate our method on piano, guitar, strings, and wind instruments, in an \textit{instrument-sensitive} (i.e., note-with-instruments, see Table~\ref{table:instrument_sensitive_v2}), or an \textit{instrument-insensitive} (see Tables~\ref{transcription_table} (piano),~\ref{table:musicnet_combined} (MusicNet test), and~\ref{guitarset_table} (GuitarSet)) manner. For the latter, only MusicNet is used, while for the former we also train an additional model using the Self-Collected data. 




For instrument-insensitive transcription (Tables~\ref{transcription_table},~\ref{guitarset_table},~\ref{table:musicnet_combined}) we report the metrics \textbf{note} (onset detection within 50ms or less) and \textbf{frame} (accuracy in detecting if a note is active/not). \textbf{Note-with-offset}, for varying thresholds, can be found in the Supplementary material. For instrument-sensitive transcription (Table~\ref{table:instrument_sensitive_v2}), we report the \textbf{note-with-instrument} metric, which uses the same 50ms timing rule, but only for notes with the correctly predicted instrument.


\subsubsection{Piano}
For piano transcription, we evaluate on the MAPS and MAESTRO test sets. 
Results can be seen in Tables~\ref{transcription_table} (instrument-insensitive) and~\ref{table:instrument_sensitive_v2} (instrment-sensitive). It can be seen that note-level accuracy is near-supervised level, even surpassing supervised-level on MAPS. This is despite training on different datasets and no direct supervision, let alone precise labeling of the exact same instrument. For frame-level accuracy, the task is more challenging, since note endings are typically weak and thus harder to decipher. While this expectedly induces lower $F1$ score for the MAESTRO dataset, we also see near-supervised performance on MAPS. Note that the same training procedure done using original MusicNet annotations yields much lower accuracy. This strongly indicates our annotation is more accurate. Similar results are achieved with self-collected data of $\sim$30 hours of piano and guitar.

\subsubsection{Guitar}
For guitar transcription, we evaluate on the GuitarSet dataset. Table~\ref{guitarset_table} demonstrates \textbf{generalization to a new instrument}, since MusicNet$_{EM}$ does not contain guitar performances. For guitar training data in Table~\ref{table:instrument_sensitive_v2} we use the self-collected $\sim5$ hours of guitar recordings together with MusicNet$_{EM}$. 
Results are consistent with the piano experiments, indicating significant improvements.

\subsubsection{String \& Wind Instruments}\label{section:invariant}
As mentioned, existing annotation of the dataset is notoriously inaccurate, and Tables~\ref{transcription_table},~\ref{guitarset_table} indicate our annotation method is more accurate. To further demonstrate this for other instruments, we evaluate on the MusicNet test set using both the original annotation and ours  
(Table~\ref{table:musicnet_combined}). Test annotation is done as described in Section~\ref{section:method}, but without the pseudo-labels step. Results can be seen in Tables~\ref{table:musicnet_combined} (instrument-insensitive) and~\ref{table:instrument_sensitive_v2} (instrument-sensitive). 

As can be seen in Table~\ref{table:musicnet_combined}, on the note-level, we have conclusive results, that our generated annotation used for training performs significantly better than training on the original annotation (over 20\% difference) on both test annotations. This indicates the method can flexibly extend to novel material with cheap labeling. 

\subsubsection{Instrument-Sensitive Transcription}\label{subsubsection:instrument_sensitive}
\begin{table*}
\caption{Instrument-sensitive Transcription results (note-with-instrument). 
We show results on the MusicNet test set, and our proposed labels -- MusicNet$_{EM}$. We also compare to~\citet{DBLP:journals/taslp/WuC020} who evaluate on the MusicNet test set.
Notice the improvements for horn, bassoon, and clarinet. 
For Violin, Cello, and Viola, accuracy according to the original annotation is comparable. 
However, this is probably due to label quality. See Supplementary material for more detail and a qualitative comparison.
We also evaluate this task on MAPS, MAESTRO, and GuitarSet. 
The most challenging instrument is viola, due to the resemblance to both violin and cello, hampering instrument identification accuracy. }\label{table:instrument_sensitive_v2}
\begin{center}
\begin{tabular}{|c|c|c|c|c|c|c|c|c|c|}
\hline
& \multicolumn{3}{|c|}{MusicNet$_{EM}$ test} & \multicolumn{3}{|c|}{MusicNet test} & \multicolumn{3}{|c|}{~\citet{DBLP:journals/taslp/WuC020}} \\
\hline
Test Set & P & R & F1 & P & R & F1 & P & R & F1 \\
\hline
MN Piano (1759, 2303, 2556, 2628) & 88.4 & 87.4 & 87.9 & 71.5 & \textbf{71.1} & \textbf{71.3} & \textbf{74.6} & 64.7 & 68.9 \\
MN Violin (2106, 2382, 2628) & 66.2 & 73.4 & 69.5 & 58.3 & 59.7 & 58.8 & \textbf{61.9} & \textbf{60.1} & \textbf{60.5} \\
MN Viola (2106, 2382) & 48.6 & 40.6 & 43.4 & \textbf{37.9} & 29.4 & \textbf{32.9} & 28.9 & \textbf{32.0} & 30.1 \\
MN Cello (2106, 2298, 2382) & 67.6 & 69.3 & 67.9 & 52.0 & \textbf{49.1} & 49.6 & \textbf{58.7} & 44.8 & \textbf{50.4} \\
MN Horn (1819, 2416) & 65.5 & 68.4 & 66.9 & \textbf{47.8} & \textbf{49.6} & \textbf{48.7} & 10.8 & 38.1 & 16.8 \\
MN Bassoon (1819, 2416) & 66.4 & 81.4 & 73.1 & \textbf{45.4} & \textbf{55.0} & \textbf{49.7} & 36.6 & 45.6. & 40.6 \\
MN Clarinet (1819, 2416) & 81.4 & 86.3 & 83.8 & \textbf{58.0} & \textbf{62.6} & \textbf{60.2} & 47.9 & 55.2 & 51.0 \\
Piano (MAESTRO) &&&& 90.5 & 76.4 & 82.3 &&& \\
Guitar (GuitarSet) &&&& 89.8 & 79.7 & 83.8 &&& \\
Piano (MAPS) &&&& 87.3 & 82.3 & 84.6 &&& \\
\hline
\end{tabular}
\end{center}
\end{table*}

\paragraph{Training \& evaluation} For Quantitative evaluation, we use the 11 instrument classes of MusicNet, with the addition of guitar, together 12 instrument classes. We evaluate on the MusicNet test set, on GuitarSet, and on MAPS. In the instrument-sensitive setting, a note is considered correct only if its predicted instrument is correct (note-with-instrument). We train on MusicNet$_{EM}$ together with the self-collected guitar data (to allow guitar detection since guitar data does not exist in the MusicNet recordings). Similar to Table~\ref{table:musicnet_combined}, we report MusicNet test results both according to our annotation, and the original annotation. Results can be seen in Table~\ref{table:instrument_sensitive_v2}. Metrics are unsurprisingly lower than Table~\ref{table:musicnet_combined}, since instrument detection is required, and confusions can occur e.g. between violin and viola. 

We believe the metrics on the original MusicNet test annotation \textbf{are far from reflecting real performance}. We provide a qualitative comparison to~\citet{DBLP:journals/taslp/WuC020} in the Video, clearly demonstrating the better performance of our approach. 
\section{Conclusion}
In this work we presented a method for multi-instrument transcription, from easily attainable unaligned supervision. We have demonstrated the method's strength for in-the-wild transcription, including cross-dataset evaluation.
We have also showed the simplicity of collecting data for our framework, which generates annotation on its own in a fully-automated process. Our work presents unprecedented transcription quality on a wide variety of instruments and genres. This work's capabilities open several new lines of research. 

Besides extending to human voices, additional effects could be added to the detection, including echo, velocity, etc. Beside the added functionality, this would probably improve the basic detection as well. In addition, adding a musical prior, driving predictions to only make sense musically (in a similar manner to a NLP) would also probably boost performance. 
Another central direction for future work is generative models. DNN based models that synthesize realistic music, although producing realistic timbre, cannot produce coherent music without conditioning on notes. Generating realistic-sounding music conditioned on notes is ideal for musicians as it enables full control over the content of the produced music. We believe the transcriptions produced using our approach can be used as a conditioning signal for training generative models, by learning the reverse mapping from transcriptions to original audio.
Finally, additional E-M iterations on small data or specific performances, even during inference, would also be an interesting avenue for future research, which we hope this work would inspire.

{\small
\bibliographystyle{IEEEtranN}
\bibliography{egbib}
}
\newpage
\appendix
\onecolumn
\section{Supplementary Material for "Unaligned Supervision for Automatic Music Transcription in The Wild"}
\subsection{Aligning real data with MIDI from a different source}
\subsubsection{Avoiding Singular Points}\label{section:labelling}
Since we align real recordings with external MIDI (i.e., from a different performer), alignment can fail at points with a contradiction in content between the two performances. This can happen when (i) one sequence has a repeated candenza while the other does not, or (ii) because of subtle nuances, and differences in precise timing of adjacent notes (e.g. in trills, or timing of individual notes within a chord). In such cases, the alignment will collapse a long segment of one sequence into a single frame in the other sequence. The long segment can be e.g. $1$ minute in case (i), or e.g. $1$ second in case (ii). Such frames that are mapped to long segments of the other sequence are called singular points. This issue is discussed by \citet{DBLP:conf/iclr/ThickstunHK17}. Their solution is to verify alignment by experts, and to exclude recordings where this occurs. This prevents the process from being fully automatic, and is less desired. Our solution is to only assign labels to non-singular points, and mask the loss from singular points. We still might assign pseudo-labels to singular points, see Subsection~\ref{subsection:pseudo} in the paper. This allows us to avoid failed alignment and also leverage all data, in a fully-automated process.

In more detail, given an audio performance with frames $1,\dots,T$, and an unaligned midi performance of the same piece with frames $1,\dots T_{target}$, the initial network predicts for each frame $1\le t\le T$ and pitch $1\le f\le 88$ probabilities for onset, frame, and offset. We denote these predictions: $P_{on}, P_{fr}, P_{off} \in[0, 1]^{T\times88}$. Similarly, we denote by $Q_{on}, Q_{fr}, Q_{off} \in\{0, 1\}^{T\times88}$ the onset, frame, and offset activations in the corresponding target midi. As local descriptors $X, Y$ for frames of the audio recording and the midi performance respectively, we use a weighted sum:
\begin{align}
&X = A * P_{on} + B* P_{fr} + C * P_{off} \\
&Y = A * Q_{on} + B* Q_{fr} + C * Q_{off}\label{Q}\\
&X\in\mathbb{R}^{T\times88},\quad Y\in\mathbb{R}^{T_{target}\times88}
\end{align}
where $A >> B >> C$, i.e., the alignment is based mainly on the \textbf{onset information}. In our experiments we used values $A = 100, B = 0.01, C = 0.001$. See Table~\ref{alignment_table} for the significant difference in accuracy, in both note- and frame-level, when aligning according to onset information, compared to aligning according to frame information.

Given a pair of sequences $X, Y$ The DTW algorithm returns an optimal alignment in the form of monotone multi-valued mappings (an index in the source can be mapped to multiple indices in the target):
\[M: X\to Y,\quad M^{-1}: Y\to X\]
where monotonicity  implies 
\[i\le j \implies k\le k'\quad\forall k\in M(i),\quad k'\in M(j).\]
and similarly for $M^{-1}$. We define the set of singular points $S = S_1\cup S_2$ where
\begin{align*}
&S_1 = \{i\,:\, |M(i)| > w\}\quad S_2 = \cup_{j: |M'(j)| > w'} M'(j)
\end{align*}
$S_1$ is the set of indices mapped to more than $w$ indices in the target domain (interval of length $>w$ in the target collapses into a single frame in the source), and $S_2$ is the set of indices mapped to indices in the target domain that cover more than $w'$ indices in the source domain (interval of length $>w'$ in the source collapses into a single frame in the target). These window sizes control a tradeoff between precision and recall. We used values $3\le w\le 9$, $w'=100$. Results in Tables~\ref{transcription_table},~\ref{table:musicnet_combined},~\ref{guitarset_table} in the paper were obtained using $w=3$, and Table~\ref{table:instrument_sensitive_v2} using $w=7$. Larger values of $w$ cause noise as they allow imprecise onset timing, and small values of $w'$ (e.g., $w'=3$) result in transcriptions that are entirely staccato.

We then assign labels to non-singular points in the following manner: Each non-singular frame $t$ in the source sequence, is mapped to a set of frames $M(t)$ in the target sequence, where $|M(t)|\le w$. We define the label $\hat{X}(t, p)$ of frame $t$ at pitch $p$ to be the maximum activation of the pitch $p$ across all frames in $M(t)$. Since we have multiple kinds of activations - onset, frame, offset, and none - we use the hierarchy: onset $>$ frame $>$ offset $>$ none.

We then assign labels only to non-singular points, in the following manner: The possible labels are: $3$ - onset, $2$ - frame, $1$ offset, and $0$ - none. We assign labels $\hat{X}$:
\begin{align}\hat{X}_t = elem\_wise\_max_{s\in M(t)}Z_s\quad i\in [T]\setminus S\label{xhat_label}\end{align}
Where $Z$ is the target label, and is defined as follows:
\[
Z = max\{3 * Q_{on}, 2 * Q_{fr}, 1 * Q_{off}\}\in [3]^{T_{target}\times 88}
\]
where $Q_{on}, Q_{fr}, Q_{off}$ are defined as in line \ref{Q} in the equation in the previous section.
Note that \[Z_s\in[3]^{88}\quad 1\le s\le T_{target}\] and the maximum over $s$ in~\ref{xhat_label} is performed entry-wise.

We back-propagate loss only from non-singular points (unless they were marked positive/negative by the pseudo-labeling which we perform afterwards). This enables us to leverage all data, and prevents the need to discard whole pieces because they contain singular points.

\subsubsection{Local-Max Adjustment}
Because of the aforementioned slight differences in precise onset timing between the real recording and its corresponding MIDI, the alignment can produce small errors in onset timing. We further refine the labels for each note independently by adjusting each note onset to be a local maximum across time (according to the predicted probabilities), which allows labeling with accurate onset timing. We do the same for note offsets. Still, offsets require further investigation since they are harder to detect.
This adjustment of onset timing is not possible when aligning spectral features of polyphonic music, as in \citet{DBLP:conf/iclr/ThickstunHK17}. A similar local-max adjustment is performed by \citet{DBLP:conf/ismir/XiBPYB18} for annotation of guitar performances, according to flux novelty (similar to spectral features) rather than a network's predicted probabilities. This however is only possible because the different guitar strings are separated, therefore the annotation is fact of monophonic music.

\section{Data \& Instrument Distribution}
\begin{table}[t]
\caption{Instrument distribution in self-collected data.}
\label{table_distrib}
\vskip 0.15in
\begin{center}
\begin{small}
\begin{sc}
\begin{tabular}{lcccr}
\toprule
Instrument & Length (Hours) \\
\midrule
Piano & 13:27:20 \\
Harpsichord & 6:20:37 \\
Harpsichord \& Strings & 3:53:21 \\
Harpsichord \& Flute & 1:02:18 \\
Guitar & 4:46:21 \\
Lute & 0:19:21 \\
Violin& 2:11:49 \\
Cello & 3:24:43 \\
Flute& 0:09:15 \\
Organ & 2:37:10 \\
Orchestra & 25:56:52 \\
Orchestra \& Piano   & 7:54:05 \\
Orchestra \& Choir & 1:49:47 \\
All & 73:52:59 \\
MusicNet & 33:43:07 \\
All, with MusicNet & 107:36:06 \\

\bottomrule
\end{tabular}
\end{sc}
\end{small}
\end{center}
\vskip -0.1in
\end{table}
As we mention in the paper, the MusicNet dataset provides recordings of multiple instruments, however, the
dataset is imbalanced. Most recordings are of solo piano ($\sim$15 out of $\sim$ 34 hours are piano solo, and $\sim$7 other hours include piano). We demonstrate the simplicity of collecting data for our method, by gathering $74$ additional hours of recordings. The full distribution of instruments can be seen in Table~\ref{table_distrib}. Transcriptions in the video are by a model trained on all data, both MusicNet and the self-collected. 

\subsection{Further Experiments \& Ablation Studies}
\label{sec:ablations}
\subsubsection{Alignment Evaluation}\label{subsection:alignment}
\begin{table*}
\caption{Alignment results. PL is short for pseudo label. Local max is the local max adjustment of onset timing.}\label{alignment_table}
\begin{center}
\begin{tabular}{|c|c|c|c|c|c|c|}
\hline
&\multicolumn{3}{|c|}{Note} & \multicolumn{3}{|c|}{Frame}\\
\hline
& P & R & F1 & P & R & F1 \\
\hline
Thresholding 0.5 & 86.2 & 83.4 & 84.7 & 76.2 & 73.6 & 74.0 \\
\hline
frame Alignment (w=3, w'=100) & 61.0 & 35.5 & 44.0 & 67.1 & 28.0 & 37.7 \\
frame Alignment + PL 0.75 (w=3, w'=50) & 85.6 & 61.3 & 70.8 & 77.3 & 56.9 & 64.2 \\
frame Alignment + PL 0.75 (w=3, w'=100) & 85.7 & 62.7 & 71.9 & 77.0 & 59.5 & 66.0 \\
\hline
onset Alignment (w=3, w'=100) & 87.7 & 83.1 & 85.2 & 75.4 & 61.5 & 66.5 \\
onset Alignment + PL 0.75 (w=1, w'=100) & \textbf{92.3} & 80.2 & 85.6 & 77.7 & 70.5 & 72.9 \\
onset Alignment + PL 0.75 (w=3, w'=100) & 91.2 & 86.8 & \textbf{88.8} & 77.2 & 76.8 & \textbf{76.1} \\
onset Alignment + PL 0.75 (w=9, w'=100) & 90.4 & 87.1 & 88.6 & 78.9 & 74.4 & 75.6 \\

onset Alignment + PL 0.5 (w=3, w'=100) & 88.3 & \textbf{87.5} & 87.7 & 74.3 & \textbf{79.5} & 75.9 \\


onset Alignment + PL 0.75 (w=3, w'=10) & 91.2 & 86.3 & 88.5 & \textbf{79.1} & 73.7 & 75.3 \\
onset Alignment + PL 0.75 (w=3, w'=50) & 91.2 & 86.7 & \textbf{88.8} & 77.5 & 76.3 & 76.0 \\

\hline
onset Alignment (local max 3) & 87.4 & 83.0 & 85.0 & 75.2 & 61.4 & 66.3\\
onset Alignment (w/o local max) & 87.9 & 82.0 & 84.7 & 75.2 & 60.8 & 66.0 \\
onset Alignment (w/o local max) + PL 0.75 & \textbf{92.3} & 84.9 & 88.3 & 77.1 & 75.4 & 75.4 \\
\hline
Thresholding (0.5) after training on the 46 pieces w/o gt labels & \textbf{92.6} & \textbf{92.8} & \textbf{92.6} & 75.3 & 77.8 & 75.3 \\

\hline

\end{tabular}
\end{center}
\end{table*}
We measure the accuracy of our labeling process on the Maestro validation dataset, for which precise annotation exists. For 46 out of the 105 pieces in the validation dataset, we were able to find additional unaligned midis (to be used instead of those offered with the dataset). We report the note and frame metrics of the alignment w.r.t the ground truth annotation, when aligment is done over precidtions on the model trained on synthetic data. We compare the results to simple thresholding. We also show the higher accuracy of aligning according to onset information rather than frame information, even for the frame-level accuracy. We show results for other parameters as well. Unless otherwise stated, we use local-max adjustment of onset timing with a window size of 7 frames. We do this in an inclusive manner: after the initial alignment, if a neighbor of an onset has a higher onset prediction, we mark it as an onset instead, and repeat this 3 times. We do this for both left and right neighbors, hence the small decrease in precision. All results can be seen in Table~\ref{alignment_table}. We also measure the accuracy on these 46 pieces, after training on them with the labels computed by the alignment (not the ground truth labels), and evaluate the accuracy of the network on them using the ground truth labels (last row in Table~\ref{alignment_table}). Main points to note in the table are: (i) Alignment according to onset information yields much more accurate annotations than aligning according to frame information, even in the frame-level metric. (ii) While annotation according to alignment alone yields slightly better annotation than thresholding with threshold 0.5, the combination of alignment, with thresholding with a higher threshold of 0.75, performs significantly better, with improvement of 4\%. (iii) The window size parameters $w, w'$ control a tradeoff between precision and recall. (iv) Local max adjustment significantly increases note-level recall, also increases frame-level recall, and gives a slight improvement in note- and frame-level F1 score. (v) The actual performance of the network on the 46 pieces after training on them with the computed annotation, is higher than the annotation's accuracy.

\subsubsection{Alignment Vs. Pseudo-Labels}
\begin{table*}
\caption{Effect of different labeling methods, evaluating on MAESTRO, MAPS, and GuitarSet datasets}\label{table:pseudo}
\begin{center}
\begin{tabular}{|c|c|c|c|c|c|c|c|c|c|c|c|c|}
\hline
 & \multicolumn{6}{|c|}{MAESTRO} & \multicolumn{6}{|c|}{MAPS} \\
\hline
 & \multicolumn{3}{|c|}{Note} & \multicolumn{3}{|c|}{Frame} & \multicolumn{3}{|c|}{Note} & \multicolumn{3}{|c|}{Frame} \\
\hline
Train method & P & R & F1 & P & R & F1 & P & R & F1 & P & R & F1 \\
\hline
Synth & 86.0 & 82.1 & 83.8 & 79.1 & 72.6 & 74.7 & 79.5 & 79.3 & 79.1 & 85.0 & 70.9 & 76.6 \\
\hline
Pseudo Labels 0.5 & 94.6 & 81.3 & 87.3 & 76.3 & 69.6 & 71.5 & 90.0 & 80.3 & 84.8 & \textbf{86.2} & 68.7 & 75.2 \\
\hline
Alignment & \textbf{95.9} & 83.0 & 88.7 & \textbf{82.6} & 63.4 & 70.5 & \textbf{90.6} & 83.2 & 86.6 & 85.8 & 66.0 & 73.7 \\
\hline
Alignment \& Pseudo Labels 0.75 & 92.6 & \textbf{87.2} & \textbf{89.7} & 77.4 &\textbf{76.1} & \textbf{76.0} & 88.2 & \textbf{86.5} & \textbf{87.3} & 84.4 & \textbf{76.7} & \textbf{79.6} \\

\hline
 & \multicolumn{6}{|c|}{GuitarSet} & \multicolumn{6}{|c|}{} \\
\hline
 & \multicolumn{3}{|c|}{Note} & \multicolumn{3}{|c|}{Frame} & \multicolumn{6}{|c|}{} \\
\hline
Train method & P & R & F1 & P & R & F1 & \multicolumn{6}{|c|}{} \\
\hline
Synth & 61.0 & \textbf{80.7} & 68.4 & 71.0 & 76.4 & 72.9 & \multicolumn{6}{|c|}{}\\
\hline
Pseudo Labels 0.5 & 81.8 & 78.6 & 79.1 & 83.4 & 73.4 & 77.4 & \multicolumn{6}{|c|}{}\\
\hline
Alignment & \textbf{90.1} & 77.4 & 82.5 & 79.2 & 80.6 & 79.4 & \multicolumn{6}{|c|}{}\\
\hline
Alignment \& Pseudo Labels 0.75 & 86.6 & 80.4 & \textbf{82.9} & \textbf{79.3} & \textbf{84.8} & \textbf{81.6} & \multicolumn{6}{|c|}{}\\
\hline
\end{tabular}
\end{center}
\end{table*}
To evaluate the contribution of each of the components - alignment with midi and pseudo-labels, we train two additional models - one where we label the real audio recordings only using pseudo-labels obtained by thresholding with a 0.5 threshold, and one where we label only using alignment. Results can be seen in Table~\ref{table:pseudo}. On the note-level, alignment alone performs better than psuedo-labels on all evaluation sets - MAPS, MAESTRO and GuitarSet. On the frame-level, alignment performs better on the MAPS test set and GuitarSet, while psuedo-labels perform better on the MAESTRO test. Our method which combines both, performs best on all three test sets, on both the note- and frame-level. 

\subsubsection{Pitch Shift}
\begin{table}
\caption{Effect of pitch shift when evaluating on MAESTRO, MAPS, and GuitarSet.}\label{table:pitch_shift}
\begin{center}
\begin{tabular}{|c|c|c|c|c|c|c|c|c|c|c|c|c|}
\hline
 & \multicolumn{6}{|c|}{MAESTRO} & \multicolumn{6}{|c|}{MAPS} \\
\hline
 & \multicolumn{3}{|c|}{Note} & \multicolumn{3}{|c|}{Frame} & \multicolumn{3}{|c|}{Note} & \multicolumn{3}{|c|}{Frame} \\
\hline
 & P & R & F1 & P & R & F1 & P & R & F1 & P & R & F1 \\
\hline
MusicNet$_{EM}$ w/o pitch shift & 91.1 & 85.6 & 88.1 & 76.3 & 74.8 & 74.3&  85.9 & 83.7 & 84.7 & 83.9 & 74.0 & 78.0 \\
\hline
MusicNet$_{EM}$ w/ pitch shift & \textbf{92.6} & \textbf{87.2} & \textbf{89.7} & \textbf{77.4} & \textbf{76.1} & \textbf{76.0} & \textbf{88.2} & \textbf{86.5} & \textbf{87.3} & \textbf{84.4} & \textbf{76.7} & \textbf{79.6} \\
\hline
 & \multicolumn{6}{|c|}{GuitarSet} & \multicolumn{6}{|c|}{} \\
\hline
& \multicolumn{3}{|c|}{Note} & \multicolumn{3}{|c|}{Frame} & \multicolumn{6}{|c|}{} \\
\hline
 & P & R & F1 & P & R & F1 & \multicolumn{6}{|c|}{} \\
\hline
MusicNet$_{EM}$ w/o pitch shift & 71.1 & \textbf{81.2} & 75.4 & 73.1 & 84.1 & 77.8 & \multicolumn{6}{|c|}{} \\
\hline
MusicNet$_{EM}$ w/ pitch shift & \textbf{86.6} & 80.4 & \textbf{82.9} & \textbf{79.3} & \textbf{84.8} & \textbf{81.6} & \multicolumn{6}{|c|}{} \\
\hline
\end{tabular}
\end{center}
\end{table}
An ablation study measuring the effect of pitch shift augmentation can be seen in Table~\ref{table:pitch_shift}: we train an additional model without pitch shift augmentation. We train both models for the same time to compensate for the smaller amount of data when training without pitch shift. For piano transcription, this augmentation gives $\sim$2\% of improvment in both note- and frame-level F1 score, increasing both precision and recall. For guitar, the improvement is 7.5\% note-level and almost 4\% frame-level. We perform the same experiment for the guitar dataset (see Section~\ref{section:experiments}). 

\subsubsection{Label Update Rate}
\begin{table*}
\caption{Effect of repeated labelling. We compare labeling once at the beginning of training, to labelling twice, to labelling 12 times at equal intervals. Best tradeoff between precision and recall is two labeling iterations. Best frame-level performance is achieved with a single labeling iteration.}\label{table_em}
\begin{center}
\begin{tabular}{|c|c|c|c|c|c|c|c|c|c|c|c|c|}
\hline
Test Set & \multicolumn{6}{|c|}{MAESTRO} & \multicolumn{6}{|c|}{MAPS} \\
\hline
Transcription Level & \multicolumn{3}{|c|}{Note} & \multicolumn{3}{|c|}{Frame} & \multicolumn{3}{|c|}{Note} & \multicolumn{3}{|c|}{Frame} \\
\hline
 & P & R & F1 & P & R & F1 & P & R & F1 & P & R & F1 \\
\hline
Synth & 86.0 & 82.1 & 83.8 & \textbf{79.1} & 72.6 & 74.7 & 79.5 & 79.3 & 79.1 & 85.0 & 70.9 & 76.6 \\
\hline
Single labelling & \textbf{95.6} & 84.7 & \textbf{89.7} & \textbf{79.1} & \textbf{76.9} & \textbf{77.0} & \textbf{90.3} & 83.7 & 86.8 & \textbf{86.2} & \textbf{78.0} & \textbf{81.4} \\
\hline
Iterative labelling (1/12) & 90.9 & 86.7 & 88.6 & 76.5 & 74.3 & 74.3 & 86.8 & 86.3 & 86.5 & 83.4 & 74.1 & 77.7 \\
\hline

Iterative labelling (1/2) & 92.6 & \textbf{87.2} & \textbf{89.7} & 77.4 & 76.1 & 76.0 & 88.2 & \textbf{86.5} & \textbf{87.3} & 84.4 & 76.7 & 79.6 \\

\hline
\end{tabular}
\end{center}
\end{table*}
To evaluate the effect of repeated updates of annotation (repeating the E-step), we train 3 models with different policies: (i) We compute the labels once only, and train on this annotation. (ii) We update the labels 12 times during training in equal intervals. (iii) We update the labels once, in the middle of training. Single labelling had the highest precision, but lower recall. Results can be seen in Table~\ref{table_em}. Policy (iii) produced the best note-level results, while policy (i) gave the best frame-level results. 

\subsubsection{Velocity}\label{section:velocity}
\begin{table}
\caption{Note with velocity results. In this metric, a note is considered correct only if its predicted velocity is within a threshold. In this metric the model trained on synthetic data performs best, velocity information does not exist for in-the-wild recordings.}\label{velocity_table}
\begin{center}
\begin{tabular}{|c|c|c|c|c|c|c|}
\hline
&\multicolumn{3}{|c|}{MAESTRO} & \multicolumn{3}{|c|}{MAPS} \\
\hline
Train Set & P & R & F1 & P & R & F1\\
\hline
Pseudo-labels & 65.2 & 61.4 & 63.2 & 63.5 & 62.4 & 62.9\\
Alignment & 56.7 & 53.5 & 55.0 & 60.5 & 59.2 & 59.8 \\
Synth & 72.2 & 69.1 & 70.5 & 66.1 & 66.3 & 65.9 \\
\hline
\end{tabular}
\end{center}

\end{table}
Dynamics and velocity are key components of any musical performance, and are a central part of the expressivity. ~\cite{DBLP:conf/ismir/HawthorneESRSRE18, DBLP:conf/iclr/HawthorneSRSHDE19} incorporate velocity into their model, i.e., the model predicts the intensity in which each note was played. The designated equipment they use for data annotation (Disklavier) also provides velocity information. However, in a weakly supervised setting such as ours, velocity becomes a challenge, since there is no direct way to recover the original note velocities from the training data, since the audio recording and the midi performance are from different sources, moreover, velocity is not necessarily well-defined. There might be some correlation between the real performances and the corresponding midi performances, but this is not guarantied. Note that velocity annotation only exists for piano datasets (MAESTRO and MAPS) but neither for GuitarSet nor MusicNet. 

When evaluating on the MAESTRO an MAPS test sets, The best velocity predictions were made by the initial model trained on synthetic data, as it was trained with full supervision over the velocity. I.e., the real data did not improve velocity prediction - see Table~\ref{velocity_table}. We tried using velocities from the midi (Table~\ref{velocity_table} AL), and using velocities predicted by the initial model as labels (Table~\ref{velocity_table} PL), but this did not improve velocity prediction. Since accurate velocity information cannot be derived from separate-source midi, we believe self-supervision is the main direction for training velocity detection, and we leave this to future work.


\subsubsection{GuitarSet Full Metrics}
\begin{table}
\caption{Transcription results on GuitarSet. MusicNet$_{EM}$ is the MusicNet recordings with our annotation. Note-level metrics of ~\citet{DBLP:conf/ismir/XiBPYB18} and~\citet{DBLP:conf/ismir/WigginsK19} are unavailable.
It is important to note that our results demonstrate generalization to a \textbf{new instrument} since the MusicNet recordings contain no guitar performances.~\citet{DBLP:journals/corr/abs-2111-03017} reach high accuracy on GuitarSet when training on GuitarSet, but perform poorly in the zero-shot task (ZS), where GuitarSet data is excluded from the train set.}\label{guitarset_table_full}
\begin{center}
\begin{tabular}{|c|c|c|c|c|c|c|}
\hline
 & \multicolumn{3}{|c|}{Note} & \multicolumn{3}{|c|}{Frame} \\
\hline
\textbf{Supervised} & P & R & F1 & P & R & F1 \\
\hline
~\citet{DBLP:conf/ismir/XiBPYB18} & - & - & - & 77.8 & 56.2 & 64.6 \\
~\citet{DBLP:conf/ismir/WigginsK19} & - & - & - & \textbf{90.0} & 76.4 & 82.6 \\
~\citet{DBLP:journals/corr/abs-2111-03017} & - & - & \textbf{90.0} & - & - & \textbf{89.0} \\
\hline
\textbf{Weakly/self-supervised} & \multicolumn{6}{|c|}{} \\
\hline
~\citet{DBLP:journals/corr/abs-2111-03017} ZS & - & - & 32.0 & - & - & 58.0 \\

MusicNet orig. & 15.0 & 8.5 & 10.0 & 71.4 & 53.3 & 57.2 \\
\hline
Synth  & 61.0 & \textbf{80.7} & 68.4 & 71.0 & 76.4 & 72.9 \\
MusicNet$_{EM}$ (ours) & 86.6 & 80.4 & \textbf{82.9} & \textbf{79.3} & \textbf{84.8} & \textbf{81.6} \\
Self-Collected (ours) & \textbf{86.7} & 79.7 & 82.2 & 75.4 & 84.7 & 79.3 \\
\hline
\end{tabular}
\end{center}
\end{table}
Results can be seen in Table~\ref{guitarset_table_full}.

\subsubsection{Frame \& Offset Detection}
\begin{table*}
\caption{Note-with-offset F1 scores for different tolerance threshold. The standard tolerance for note-with-offset is the maximum between 50ms and 20\% of reference note length. We show results also for higher tolerance as follows: we increase the tolerance to 250, 500, 1000, and 2000ms, keeping the 20\% threshold fixed (rows 4-7), and increase the tolerance to 40, 50, 100, 200, 300\%, keeping the 50ms threshold fixed (rows 8-12). For low tolerance, results are inconclusive between the model trained on synthetic data, our method, and pseudo-labels. As can be expected, as the tolerance increases, the note-with-offset F1 score becomes closer to the note-level F1 score, and when reaching a 0.5s tolerance (rows 5-7), our method achieves highest note-with-offset F1 score on all three test sets. .}\label{table:offset}
\begin{center}
\begin{tabular}{|c|c|c|c|c|c|c|c|c|c|c|c|c|}
\hline
& \multicolumn{4}{|c|}{MAPS} & \multicolumn{4}{|c|}{MAESTRO} & \multicolumn{4}{|c|}{GuitarSet} \\
\hline
Threshold (s, \%) & Synth & Ours & PL & Sup. & Synth & Ours & PL & Sup. & Synth & Ours & PL & Sup.\\
\hline
0.05, 20 (def.) & 42.5 & 52.2 & 46.6 & 67.4 & 43.6 & 39.6 & 39.7 & 80.3 & 35.7 & 48.8 & 35.6 & 78.0 \\
\hline
0.25, 20 & 57.3 & 66.9 & 60.9 & - & 54.6 & 56.2 & 52.5 & 83.1  & 58.2 & 67.0 & 59.6 & 86.0 \\
0.5, 20 & 65.4 & 73.5 & 68.9 & - & 62.9 & 66.3 & 61.5 & 85.5 & 62.1 & 71.9 & 63.8 & 90.0 \\
1.0, 20 & 72.0 & 78.9 & 75.2 & - & 71.2 & 75.3 & 70.4 & 88.8 & 65.2 & 75.7 & 66.8 & - \\
2.0, 20 & 75.9 & 82.5 & 79.1 & - & 77.1 & 81.7 & 76.5 & 91.3 & 67.2 & 78.1 & 68.3 & - \\
\hline
0.05, 40 & 46.0 & 58.7 & 49.9 & - & 49.4 & 47.0 & 46.7 & 82.6 & 50.9 & 61.4 & 50.8 & - \\
0.05, 50 & 48.7 & 62.3 & 52.5 & - & 52.3 & 50.2 & 49.9 & 83.8 & 55.4 & 64.7 & 55.9 & - \\
0.05, 100 & 61.0 & 77.2 & 64.4 & - & 68.7 & 66.0 & 67.5 & 89.9 & 62.2 & 70.4 & 64.4 & - \\
0.05, 200 & 67.8 & 80.9 & 71.1 & - & 73.6 & 72.8 & 72.8 & 91.3 & 64.5 & 74.3 & 66.3 & - \\
0.05, 300 & 71.5 & 82.5 & 74.6 & - & 76.3 & 76.9 & 75.7 & 92.0 & 65.7 & 76.1 & 67.4& - \\
\hline
\end{tabular}
\end{center}
\end{table*}
Onsets by definition are the initial appearance, or beginning of notes, and their lengths do not vary between notes - long notes and short notes have an onset with the same length, which is typically defined to be a single frame. Thus, there is a strict correspondence between onsets in a real performance and its corresponding midi, up to a warping function. However, frame activation determines the duration of a note, which lasts several frames and can significantly vary between different notes. The musical score of a piece has instructions for note duration, which provides approximate information that enables learning frame-level transcription in the weakly supervised setting. However, small discrepancies can exist between the real and the midi performances, even after warping, as the exact time of offset can slightly vary between performances. Therefore, although there is improvement in frame-level accuracy gained through weak supervision, it is moderate. These small discrepancies in performance explain the gap between supervised and weakly supervised learning in the frame-level accuracy in Table~\ref{transcription_table} (79.6-81.4\% vs. 84.9\%) and between note-level accuracy and frame-level accuracy in the weakly supervised setting (79.6-81.4\% vs. 87.3\%). However, as we've explained in Section~\ref{section:introduction}, the human ear is sensitive mainly to the onset time, and less to the notes' precise duration and offset time, assuming note duration is approximately correct.

To measure the accuracy of our trained model in detecting note offsets, we compute the note-with-offset level metrics for different thresholds. The standard tolerance for offset detection is 50 milliseconds, or \%20 of the note length, whichever is greater. Results can be seen in Table~\ref{table:offset}. It can be seen that the contribution of unaligned supervision to offset detection is small, and increases as the offset tolerance thresholds are increased.

We believe frame-level detection, together with offset detection, can be further improved through time-stretching consistency, and this is an important direction for future work.

\subsubsection{MAESTRO with unaligned supervision}
\begin{table}
\caption{Training on MAESTRO with unaligned supervision. For $\sim$7 hours of the MAESTRO validation set, we find unaligned MIDI of the same pieces from unrelated performers, and denote this data MAESTRO$_{EM}$. First row - accuracy when training on MAESTRO$_{EM}$ and evaluating on MAESTRO$_{EM}$, but w.r.t. the GT labels. Second row - training on both MAESTRO$_{EM}$ and MusicNet$_{EM}$, and evaluating on the MAESTRO test set. Metrics in row 3 from~\citet{DBLP:conf/iclr/HawthorneSRSHDE19}. Notice the small gap in note-level metrics between rows 1 (unaligned supervision) and 3 (full supervision), while the training scheme in row 1 is applicable to any instrument.}\label{train_table}
\begin{center}
\begin{tabular}{|c|c|c|c|c|c|c|c|}
\hline
&\multicolumn{3}{|c|}{Note} & \multicolumn{3}{|c|}{Frame} \\
\hline
& P & R & F1 & P & R & F1 \\
\hline
\textbf{train MAESTRO$_{EM}$, test: MAESTRO$_{EM}$ GT} & 95.2 & 90.4 & 92.7 & 78.1 & 77.7 & 77.2 \\
\textbf{train: MusicNet$_{EM}$ + MAESTRO$_{EM}$, test: MAESTRO test} 
& 93.9 & 88.6 & 91.1 & 72.3 & 85.4 & 78.0 \\
MAESTRO train acc. (Supervised) & 98.9 & 94.4 & 96.6 & 94.2 & 92.6 & 93.4 \\

train: Synth, test: MAESTRO$_{EM}$ GT & 86.2 & 83.6 & 84.8 & 76.5 & 74. & 74.3 \\
train: MusicNet$_{EM}$, test: MAESTRO$_{EM}$ GT & 93.3 & 88.6 & 90.8 & 77.6 & 74.5 & 75.5 \\

\hline
\end{tabular}
\end{center}
\end{table}
An important question that arises is what is the accuracy on the test set, when some samples from the test domain, or samples similar to the test domain, are seen during training, but without labels, only unaligned supervision. To evaluate this, we searched for midi performances of pieces in the MAESTRO dataset, unaligned and by other performers. We were able to find such performances for 46 pieces from the MAESTRO validation set, of total time 6:57:22. We denote this by MAESTRO$_{EM}$. We conduct two experiments: (i) We train on MAESTRO$_{EM}$ alone using our method, without the ground truth labels, and then measure accuracy on MAESTRO$_{EM}$ w.r.t. the ground truth labels. (ii) In another experiment, we add MAESTRO$_{EM}$ to MusicNet$_{EM}$ to measure the effect on the MAESTRO test set. Results can be seen in Table~\ref{train_table}, rows 1-2.

\end{document}